\documentclass[preprint,review,12pt]{elsarticle}
\usepackage[english]{babel}
\usepackage{url} 
\usepackage{hyperref}
\usepackage{amssymb}
\usepackage{amsfonts}
\usepackage{amstext}
\usepackage{amsmath}
\usepackage{slashed} 
\usepackage{graphicx} 
\usepackage{color}
\usepackage{bm} 
\newcommand{\corr}[1]{\textcolor{black}{#1}}

\begin{document}
\allowdisplaybreaks

\title{Analytical Infrared Limit of Nonlinear Thomson Scattering Including Radiation Reaction}

\author{A. Di Piazza}
\address{Max Planck Institute for Nuclear Physics, Saupfercheckweg 1, D-69117 Heidelberg, Germany}

\begin{abstract}
If an electric charge is accelerated by a sufficiently intense electromagnetic field, the effects of the radiation emitted by the charge on the charge dynamics (radiation reaction) cannot be ignored. Here we show that classical radiation-reaction effects alter qualitatively and quantitatively the infrared behavior of the spectrum of the radiation emitted by an electron in the presence of an intense electromagnetic plane wave (nonlinear Thomson scattering). An analytical expression of the infrared limit of nonlinear Thomson scattering is provided, which includes radiation-reaction effects and is valid for an arbitrary plane wave. Apart from their own conceptual importance and as a signature of classical radiation reaction, these results provide the limiting expression of the corresponding and yet unknown exact infrared behavior of strong-field QED in an intense plane wave.

\vspace{1cm}

\end{abstract}

\maketitle

\section{Introduction}
The definitive foundation of classical electrodynamics culminated with the formulation of Maxwell's equations, which, together with the Lorentz equation, allow in principle to describe self-consistently the dynamics of electric charges and their electromagnetic field. Yet the full description of the coupled dynamics of a \emph{single} elementary charge, an electron for definiteness, and of its own or ``self'' electromagnetic field in the presence of an external force has revealed one of the most profound physical inconsistencies of classical electrodynamics: the problem of the electron self-energy. Indeed, the inclusion of the ``reaction'' of the self electromagnetic field on the electron dynamics (radiation reaction) has to confront an unavoidable Coulomb-like divergence when evaluating the self field at the electron position \cite{Jackson_b_1975,Landau_b_2_1975,Barut_b_1980,Rohrlich_b_2007}. After absorbing the divergent self electromagnetic energy via a redefinition of the electron mass, the resulting equation of motion of the electron features an additional, finite ``radiation-reaction'' force and it is known as Lorentz-Abraham-Dirac (LAD) equation \cite{Abraham_b_1905,Lorentz_b_1909,Dirac_1938}. In the case of interest here, where the external force is also electromagnetic, the LAD equation is derived by eliminating from the Maxwell-Lorentz system of equations the electromagnetic field generated by the electron. In other words, solving the LAD equation amounts to solving exactly the electron dynamics in the external electromagnetic field and plugging the resulting solution into the Li\'{e}nard-Wiechert potentials amounts to determining the corresponding exact electromagnetic field.

Now, it is known that the LAD equation has unexpected features because the radiation-reaction force contains the electron acceleration and its time-derivative. Moreover, it is plagued by serious physical inconsistencies like the allowance of the so-called runaway solutions, where the electron's acceleration exponentially increases with time even if the external field, for example, vanishes identically \cite{Jackson_b_1975,Landau_b_2_1975,Barut_b_1980,Rohrlich_b_2007}. The origin of the existence of the runaway solutions is precisely a term in the radiation-reaction force proportional to the time-derivative of the electron acceleration. Landau and Lifshitz realized that within the realm of classical electrodynamics, i.e., if quantum effects are negligible, the radiation-reaction force in the instantaneous rest frame of the electron is always much smaller than the Lorentz force \cite{Landau_b_2_1975}. This allows one to replace the electron four-acceleration in the radiation-reaction four-force with its ``zero-order'' expression, given by the Lorentz four-force divided by the electron mass \cite{Landau_b_2_1975}. The conceptual importance of the ``reduction of order'' put forward by Landau and Lifshitz is that the neglected quantities are much smaller than quantum effects, which are already ignored classically. The resulting equation, known as Landau-Lifshitz (LL) equation, turns out to be free of the physical inconsistencies of the LAD equation. The equivalence between the LL equation and the LAD equation within the realm of classical electrodynamics\corr{, in the sense that they differ by terms much smaller than quantum corrections,} has been confirmed numerically in \cite{Koga_2004} in the case of a plane-wave background field and, numerically and analytically in \cite{Bulanov_2011} for other non-plane-wave electromagnetic background fields. Instead, the equivalence between the LL equation and the Ford-O'Connell equation, where the reduction of order is carried out in a different but classically equivalent way as for the LL equation, has been confirmed numerically in \cite{Kravets_2013}. Presently the LL equation and in general the problem of radiation reaction are under active investigation both theoretically \cite{Vranic_2014,Blackburn_2014,Tamburini_2014,Li_2014,Heinzl_2015,Yoffe_2015,Capdessus_2015,Vranic_2016,Dinu_2016,Di_Piazza_2017,Harvey_2017,Ridgers_2017,Niel_2017,Niel_2018} and experimentally \cite{Wistisen_2018,Cole_2018,Poder_2017} (see the recent reviews \cite{Hammond_2010,Di_Piazza_2012,Burton_2014} for previous publications).

In the present Letter we focus on the established equivalence between solving the LAD/LL equation for an electron in an external electromagnetic field and solving the coupled Maxwell's and Lorentz equations, i.e., determining the exact electron's dynamics in that field (in the case of the LL equation, short of effects much smaller than quantum effects). By exploiting this idea, we derive analytically the classical infrared limit of the energy spectrum emitted by an electron driven by an arbitrary plane wave (nonlinear Thomson scattering) including radiation-reaction effects. It is known, in fact, that according to the Lorentz equation, the asymptotic momentum of the electron after exiting an arbitrary plane wave with no dc component coincides with the asymptotic one before the electron enters the plane wave (Lawson-Woodard theorem \cite{Woodward_1946,Woodward_1948}). This implies that the infrared limit of the spectrum of nonlinear Thomson scattering vanishes. Below we show that the situation qualitatively changes if the classical dynamics of the electron in the plane wave is determined according to the LL equation, i.e., by including the effects of the electron self field. The exact analytical solution of the LL equation in a plane wave \cite{Di_Piazza_2008_a}, in fact, shows that the two asymptotic momenta are different, which alters the infrared asymptotic behavior of the emitted energy spectrum via nonlinear Thomson scattering. It is worth pointing out the difference with respect to the results found in \cite{Dinu_2012,Ilderton_2013b}, where the authors investigate both classically and quantum mechanically the infrared behavior of the electron emission spectrum in a plane wave. Interestingly, the findings in \cite{Dinu_2012,Ilderton_2013b} show that the asymptotic initial and final momenta of the electron can be different already within the Lorentz dynamics if the plane wave has a dc component. The quantum counterpart of this effect is also investigated in \cite{Dinu_2012,Ilderton_2013b} and the quantum results are shown to be consistent with the classical ones. \corr{We} also mention here that the effects of radiation reaction on the energy spectrum emitted by an electron in the presence of a time-dependent electric field of the form $E(t)=k\delta(t)$, with $k$ being a constant, have been investigated by Dirac in \cite{Dirac_1938}.

Unless otherwise explicitly stated, units with $\hbar=c=4\pi\epsilon_0=1$ are employed throughout. The metric tensor is $\eta^{\mu\nu}=\text{diag}(+1,-1,-1,-1)$.

\section{Infrared behavior of classical radiation}

Let us consider an electron (charge $e<0$ and mass $m$, respectively), whose trajectory is characterized by the instantaneous position $\bm{r}=\bm{r}(t)$, the instantaneous velocity $\bm{\beta}=\bm{\beta}(t)$, and the instantaneous acceleration $\dot{\bm{\beta}}=\dot{\bm{\beta}}(t)$. The electromagnetic energy $\mathcal{E}$ radiated by the electron per unit of angular frequency $\omega$ and along the direction $\bm{n}=(\sin\vartheta_n\cos\varphi_n,\sin\vartheta_n\sin\varphi_n,\cos\vartheta_n)$ within a solid angle $d\Omega_n=\sin\vartheta_n d\vartheta_n d\varphi_n$ is given by  [see, e.g., Eq. (14.65) in \cite{Jackson_b_1975}]:
\begin{equation}
\label{dE_dodO}
\frac{d\mathcal{E}}{d\omega d\Omega_n}=\frac{e^2}{4\pi^2}\left|\int_{-\infty}^{\infty} dt\frac{\bm{n}\times[(\bm{n}-\bm{\beta})\times\dot{\bm{\beta}}]}{(1-\bm{n}\cdot\bm{\beta})^2}e^{i\omega(t-\bm{n}\cdot\bm{r})}\right|^2.
\end{equation}
Since the pre-exponential factor in the integrand in Eq. (\ref{dE_dodO}) is equal to the time derivative of $\bm{n}\times(\bm{n}\times\bm{\beta})/(1-\bm{n}\cdot\bm{\beta})$, it is clear that the infrared limit $\omega\to 0$ of Eq. (\ref{dE_dodO}) reads \cite{Jackson_b_1975}
\begin{equation}
\label{dE_dodO_0}
\left.\frac{d\mathcal{E}}{d\omega d\Omega_n}\right|_{\omega\to 0}=\frac{e^2}{4\pi^2}\left[\frac{\bm{n}\times(\bm{n}\times\bm{\beta}_f)}{1-\bm{n}\cdot\bm{\beta}_f}-\frac{\bm{n}\times(\bm{n}\times\bm{\beta}_i)}{1-\bm{n}\cdot\bm{\beta}_i}\right]^2,
\end{equation}
where $\bm{\beta}_{f/i}=\bm{\beta}(\pm\infty)$. By introducing the corresponding asymptotic four-momenta $p^{\mu}_{f/i}=(\varepsilon_{f/i},\bm{p}_{f/i})$, with $\varepsilon_{f/i}=m\gamma_{f/i}=m/(1-\beta^2_{f/i})^{1/2}$ and $\bm{p}_{f/i}=\varepsilon_{f/i}\bm{\beta}_{f/i}$, and the four-dimensional quantity $n^{\mu}=(1,\bm{n})$, one can easily show that Eq. (\ref{dE_dodO_0}) can be rewritten in the covariant-like form \cite{Peskin_b_1995}
\begin{equation}
\label{dE_dodO_0_f}
\left.\frac{d\mathcal{E}}{d\omega d\Omega_n}\right|_{\omega\to 0}=\frac{e^2}{4\pi^2}\left[\frac{2(p_ip_f)}{(np_i)(np_f)}-\frac{m^2}{(np_i)^2}-\frac{m^2}{(np_f)^2}\right],
\end{equation}
where the on-shell conditions $p_{f/i}^2=m^2$ have been employed. Before specializing to the case of a plane wave, we observe that the integral of Eq. (\ref{dE_dodO_0_f}) over the solid angle can be taken exactly for arbitrary asymptotic four-momenta $p^{\mu}_{f/i}$. In fact, since the integral is invariant under spatial rotations, one can always assume that $\bm{p}_i$ lies along the $z$ axis and that $\bm{p}_f$ lies on the $x\text{-}z$ plane. The final result is
\begin{equation}
\label{dE_do}
\left.\frac{d\mathcal{E}}{d\omega}\right|_{\omega\to 0}=\frac{2}{\pi}e^2\left[\frac{\rho}{\sqrt{\rho^2-1}}\log\big(\rho+\sqrt{\rho^2-1}\big)-1\right],
\end{equation}
which is a function of the Lorentz-invariant quantity $\rho=(p_ip_f)/m^2\ge 1$. This also implies that one could have used Lorentz invariance to evaluate the integral: by imagining to work in the initial rest-frame of the electron and by indicating there the physical quantities with a prime ($p_i^{\prime\,\mu}=(m,\bm{0})$), the integral in $d\Omega'_n$ is easily taken and one would obtain Eq. (\ref{dE_do}), with $\rho=\varepsilon'_f/m$. Since the function in Eq. (\ref{dE_do}) is ``universal'', i.e., independent of the specific problem at hand, and it essentially depends only on one physical quantity, it is worth plotting it (see Fig. 1). Also, for the sake of completeness, we report its two limiting expressions $d\mathcal{E}/d\omega|_{\omega\to 0}\approx (4/3\pi)e^2(\rho-1)$, for $\rho\to 1^+$, and $d\mathcal{E}/d\omega|_{\omega\to 0}\approx (2/\pi)e^2[\log(2\rho)-1]$, for $\rho\to\infty$.
\begin{figure}
\begin{center}
\includegraphics[width=\columnwidth]{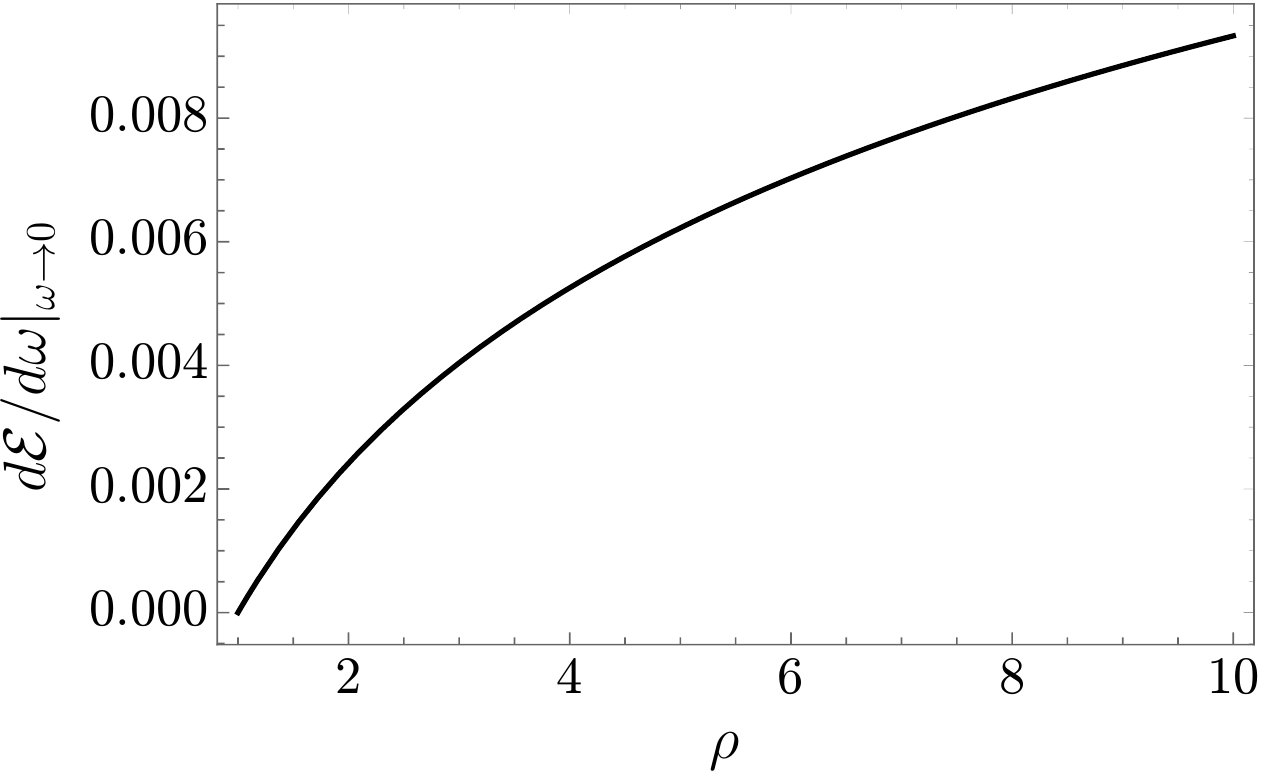}
\caption{Differential emitted energy $d\mathcal{E}/d\omega|_{\omega\to 0}$ as a function of $\rho=(p_ip_f)/m^2$.}
\end{center}
\end{figure}
It is clear that the expression $d\mathcal{E}/d\omega|_{\omega\to 0}$ cannot be further integrated over $\omega$ to obtain the total emitted energy $\mathcal{E}$. However, to this end the relativistic Larmor formula \cite{Jackson_b_1975}
\begin{equation}
\label{Larmor}
\frac{d\mathcal{E}}{dt}=-\frac{2}{3}e^2\frac{du^{\mu}}{ds}\frac{du_{\mu}}{ds}, 
\end{equation}
where $u^{\mu}(s)=p^{\mu}(s)/m=(\varepsilon(s),\bm{p}(s))/m$ is the electron four-velocity and $s$ its proper time, can be employed (see also below).

\section{Analytical infrared limit of nonlinear Thomson scattering}

Now, we consider a plane-wave background field, described by the four-vector potential $A^{\mu}(\phi)=(A^0(\phi),\bm{A}(\phi))$, where $\phi=(n_0x)=t-\bm{n}_0\cdot\bm{x}$, with $n_0^{\mu}=(1,\bm{n}_0)$ and the unit vector $\bm{n}_0$ identifying the propagation direction of the plane wave itself. We work in the Lorenz gauge $\partial_{\mu}A^{\mu}(\phi)=(n_0A'(\phi))=0$ with the additional condition $A^0(\phi)=0$. Here and below the prime indicates the derivative with respect to the argument of a function. By assuming that $\lim_{\phi\to\pm\infty}\bm{A}(\phi)=0$ (i.e., that the plane wave has no dc component), then the Lorenz-gauge condition implies $\bm{n}_0\cdot\bm{A}(\phi)=0$. Thus, the four-vector potential $A^{\mu}(\phi)$ can be written as $A^{\mu}(\phi)=\sum_{j=1}^2a_j^{\mu}\psi_j(\phi)$, where the four-vectors $a_j^{\mu}$ have the form $a_j^{\mu}=(0,\bm{a}_j)$ and fulfill the conditions $(a_ja_{j'})=-\bm{a}_j^2\delta_{jj'}$, with $j,j'=1,2$, and $(n_0a_j)=-\bm{n}_0\cdot\bm{a}_j=0$, and where the functions $\psi_j(\phi)$ are arbitrary (physically well-behaved) functions such that $\lim_{\phi\to\pm\infty}\psi_j(\phi)=0$. From the exact solution of the LL equation in an arbitrary plane wave \cite{Di_Piazza_2008_a} (see the Appendix A) it can be seen that the asymptotic final electron four-momentum $p_f^{\mu}$ is given by
\begin{equation}
\label{iSol}
\begin{split}
p_f^{\mu}=&\frac{1}{h_f}\left[p_i^{\mu}+\frac{1}{2\eta_0}(h^2_f-1)k_0^{\mu}+\frac{1}{\eta_0}\left(\mathcal{I}_{1,f}\frac{eF_1^{\mu\nu}}{m^2}+\mathcal{I}_{2,f}\frac{eF_2^{\mu\nu}}{m^2}\right)p_{i,\nu}\right.\\
&\left.+\frac{1}{2\eta_0}(\xi_1^2\mathcal{I}^2_{1,f}+\xi_2^2\mathcal{I}^2_{2,f})k_0^{\mu}\right],
\end{split}
\end{equation}
where
\begin{align}
\label{h_f}
h_f&=1+\frac{2}{3}e^2\eta_0\int_{-\infty}^{\infty}d\varphi[\xi_1^2\psi^{\prime\,2}_1(\varphi)+\xi_2^2\psi^{\prime\,2}_2(\varphi)],\\
\label{I_jf}
\mathcal{I}_{j,f}&=-\frac{2}{3}e^2\eta_0\int_{-\infty}^{\infty}d\varphi\,\psi_j(\varphi)[\xi_1^2\psi^{\prime\,2}_1(\varphi)+\xi_2^2\psi^{\prime\,2}_2(\varphi)].
\end{align}
In the above equations, we have introduced the normalized field amplitudes $\xi_j=|e\bm{a}_j|/m$, the central four-wave-vector $k_0^{\mu}=\omega_0n_0^{\mu}$ of the plane wave, with $\omega_0$ being its central angular frequency, and the related quantities $\eta_0=(k_0p_i)/m^2$ and $F_j^{\mu\nu}=k_0^{\mu}a_j^{\nu}-k_0^{\nu}a_j^{\mu}$, whereas the time dependence of the physical quantities has been expressed via the plane-wave phase $\varphi=(k_0x)$. The corresponding solution of the Lorentz equation is obtained by formally setting $e^2=0$ in $h_f$ and in $\mathcal{I}_{j,f}$ (see also the Appendix A) and shows that $p^{\mu}_f=p^{\mu}_i$ in this case, according to the Lawson-Woodard theorem mentioned in the introduction. The equality $p^{\mu}_f=p^{\mu}_i$ does not hold anymore once radiation-reaction effects are included and the fact, in particular, that $p_{f,-}\neq p_{i,-}$ has been indicated in \cite{Harvey_2011b} as a possible signature to measure radiation-reaction effects. In other words, unlike the leading-order dynamics of the electron determined exclusively by the external plane wave, the electron dynamics, which includes the effects of the self electromagnetic field, induces an overall change in the electron four-momentum. This in turn implies that in general the quantity $\rho=(p_ip_f)/m^2$ is larger than unity. In fact, it can easily be shown to be given by
\begin{equation}
\label{rho}
\rho=\frac{1}{2}\left(h_f+\frac{1}{h_f}\right)+\frac{\xi_1^2\mathcal{I}^2_{1,f}+\xi_2^2\mathcal{I}^2_{2,f}}{2h_f}.
\end{equation}
If one imagines that radiation-reaction effects, i.e., higher-order radiative effects, induce small corrections to the Lorentz dynamics, Eq. (\ref{rho}) indicates two different scaling laws of these effects. On the one hand, the leading correction resulting from the function $h_f$ scales with the square of the parameter $R_C=e^2\eta_0\xi_0^2$ (see also \cite{Di_Piazza_2008_a}), where $\xi_0=\sqrt{\xi_1^2+\xi_2^2}$ gives a measure of the strength of the plane wave \cite{Ritus_1985,Di_Piazza_2012}. On the other hand, the corrections resulting from the functions $\mathcal{I}_{j,f}$ scale with the square of the parameter $R_C\xi_0=e^2\eta_0\xi_0^3$. This is potentially highly beneficial because available high-power laser systems can routinely exceed the threshold $\xi_0=1$ \cite{Di_Piazza_2012}. However, as it is known, the plane-wave pulse duration also plays a role here, which we study below by means of two paradigmatic examples. First, we consider the case of a linearly polarized pulsed plane wave characterized by $\psi_2(\varphi)=0$ and by
\begin{equation}
\psi_1(\varphi)=\psi_{G_1}(\varphi)=e^{-\varphi^2/2\Phi^2}\sin(\varphi+\varphi_0),
\end{equation}
where the parameter $\Phi$ gives a measure of the pulse duration and the parameter $\varphi_0$ is the so-called carrier envelope phase. Both quantities $h_f$ and $\mathcal{I}_{1,f}$ can be calculated exactly and, by denoting them as $h_{G_1,f}$ and $\mathcal{I}_{G_1,f}$, respectively, are given by
\begin{align}
\label{h_f_G}
h_{G_1,f}=&1+\frac{\sqrt{\pi}}{3}R_C\Phi\left\{1+\frac{1}{2\Phi^2}\left[1-\cos(2\varphi_0)e^{-\Phi^2}\right]\right\},\\
\label{I_f_G}
\mathcal{I}_{G_1,f}=&-\sqrt{\frac{8\pi}{243}}R_C\Phi e^{-\Phi^2/6}\left[\sin(\varphi_0)\left(1+\frac{3}{4\Phi^2}\right)-\frac{1}{4\Phi^2}\sin(3\varphi_0)e^{-4\Phi^2/3}\right].
\end{align}
Since even for a single-cycle pulse it is $\Phi\gtrsim 2\pi$, for the sake of simplicity we can neglect the correcting terms proportional to $1/\Phi^2\lesssim 2.5\times 10^{-2}$ in Eqs. (\ref{h_f_G})-(\ref{I_f_G}) and we obtain the following approximated expression of $\rho_{G_1}$:
\begin{equation}
\label{rho_G1}
\begin{split}
\rho_{G_1}\approx& \frac{1}{2}\left[\left(1+\frac{\sqrt{\pi}}{3}R_C\Phi\right)+\left(1+\frac{\sqrt{\pi}}{3}R_C\Phi\right)^{-1}\right]\\
&+\frac{4\pi}{243}\sin^2(\varphi_0)\frac{R^2_C\xi_0^2\Phi^2e^{-\Phi^2/3}}{1+(\sqrt{\pi}/3)R_C\Phi}.
\end{split}
\end{equation}
Note that, in general, due to the different scaling with respect to $\xi_0^2$, the term arising from $\mathcal{I}_{G_1,f}$ [the one on the second line in Eq. (\ref{rho_G1})] cannot be neglected. However, if one takes into account that quantum effects can be neglected only if $\eta_0\ll 1$ and $\chi_0=\eta_0\xi_0\ll 1$ (see, e.g., \cite{Landau_b_2_1975,Ritus_1985}), that term can indeed be safely neglected even for upcoming high-power laser systems because of the expected pulse durations such that $\Phi\gtrsim 10\pi$ \cite{ELI,Apollon_10P,XCELS}. From now on we assume that this is indeed the case. The quantity $d\mathcal{E}_{G_1}/d\omega d\Omega_n|_{\omega\to 0}$ is obtained in general by substituting the resulting approximated expression of $\rho_{G_1}$ in Eq. (\ref{dE_do}). For the sake of completeness, we report the asymptotic expressions of $d\mathcal{E}_{G_1}/d\omega|_{\omega\to 0}$ in the two cases $R_C\Phi\ll 1$ and $R_C\Phi\gg 1$:
\begin{equation}
\label{dE_do_G1}
\left.\frac{d\mathcal{E}_{G_1}}{d\omega}\right|_{\omega\to 0}\approx
\begin{cases}
\frac{2}{27}e^2R_C^2\Phi^2 & R_C\Phi\ll 1,\\
\frac{2}{\pi}e^2\left[\log\left(\frac{\sqrt{\pi}}{3}R_C\Phi\right)-1\right] & R_C\Phi\gg 1.
\end{cases}
\end{equation}

Now, the suppressing exponential factor in $\mathcal{I}_{G_1,f}$ discussed above [see Eq. (\ref{rho_G1})] ultimately arises from the fact that the function $\psi_{G_1}(\varphi)\psi^{\prime\,2}_{G_1}(\varphi)$ is oscillating [see Eq. (\ref{I_jf})]. Thus, in order to overcome this drawback, we follow \cite{Tamburini_2014} and use a linearly-polarized, two-color pulse:
\begin{equation}
\psi_{G_2}(\varphi)=\frac{e^{-\varphi^2/2\Phi^2}}{\sqrt{1+4\zeta^2}}[\sin(\varphi+\varphi_0)+\zeta\sin(2\varphi+\varphi_{\zeta})],
\end{equation}
which also employs the second harmonics of the fundamental driving field and which in general allows for the function $\psi_{G_2}(\varphi)\psi^{\prime\,2}_{G_2}(\varphi)$ to feature a dc component. Here, the non-negative parameter $\zeta$ gives a measure of the relative amplitude of the two components and the constant $\varphi_{\zeta}$ accounts for a possible phase shift between the two components. In order to make a fair comparison with the single-Gaussian case, we have rescaled the two-color field by the factor $\sqrt{1+4\zeta^2}$ in such a way that the average intensity of the two fields is the same (strictly speaking for $\Phi\gg 1$).

By keeping again only the leading terms in the parameter $\Phi$, we obtain in this two-color-Gaussian case
\begin{equation}
\label{rho_G2}
\begin{split}
\rho_{G_2}\approx& \frac{1}{2}\left[\left(1+\frac{\sqrt{\pi}}{3}R_C\Phi\right)+\left(1+\frac{\sqrt{\pi}}{3}R_C\Phi\right)^{-1}\right]\\
&+\frac{\pi}{12}\frac{\zeta^2\sin^2(2\varphi_0-\varphi_{\zeta})}{(1+4\zeta^2)^3}\frac{R^2_C\xi_0^2\Phi^2}{1+(\sqrt{\pi}/3)R_C\Phi}.
\end{split}
\end{equation}
The second line of this result indeed shows the theoretical possibility of fully exploiting the additional factor $\xi_0^2$ with an appropriate choice of $\zeta$ and $\varphi_{\zeta}$. Indeed, knowing $\varphi_0$, one can always choose $\varphi_{\zeta}$ such that $|\sin(2\varphi_0-\varphi_{\zeta})|=1$, whereas the maximum of the function $\zeta^2/(1+4\zeta^2)^3$ is $1/27$ at $\zeta^2=1/8$. In this respect, it is convenient to set $g^2/27=\sin^2(2\varphi_0-\varphi_{\zeta})\zeta^2/(1+4\zeta^2)^3$ and to keep in mind that $g^2$ can be chosen to be approximately unity. As for the case of the single-color Gaussian pulse, the general expression of $d\mathcal{E}_{G_2}/d\omega d\Omega_n|_{\omega\to 0}$ is obtained by replacing Eq. (\ref{rho_G2}) in Eq. (\ref{dE_do}). Here, it is interesting to report the two asymptotic expressions of $\rho_{G_2}$ for $R_C\Phi\ll 1$ and $R_C\Phi\gg 1$ under the experimentally relevant conditions $g\sim 1$ and $\xi_0\gg 1$: 
\begin{equation}
\rho_{G_2}\approx
\begin{cases}
1+\frac{\pi}{324}g^2R_C^2\Phi^2\xi_0^2 & R_C\Phi\ll 1,\\
\frac{\sqrt{\pi}}{108}g^2R_C^2\Phi^2\xi_0^2 & R_C\Phi\gg 1.
\end{cases}
\end{equation}
Unlike for the single-color Gaussian pulse, in the case $R_C\Phi\ll 1$ one has to employ the general expression of $d\mathcal{E}/d\omega d\Omega_n|_{\omega\to 0}$ in Eq. (\ref{dE_do}) if $R_C\Phi\xi_0\gtrsim 1$. For the sake of completeness, we report here the asymptotic expressions of $d\mathcal{E}_{G_2}/d\omega|_{\omega\to 0}$ for $R_C\Phi\xi_0\ll 1$ and for $R_C\Phi\gg 1$:
\begin{equation}
\label{dE_do_G2}
\left.\frac{d\mathcal{E}_{G_2}}{d\omega}\right|_{\omega\to 0}\approx
\begin{cases}
\frac{1}{243}e^2g^2R_C^2\Phi^2\xi_0^2 & R_C\Phi\xi_0\ll 1,\\
\frac{2}{\pi}e^2\left[\log\left(\frac{\sqrt{\pi}}{54}g^2R_C\Phi\xi^2_0\right)-1\right] & R_C\Phi\gg 1.
\end{cases}
\end{equation}

In order to have an idea of the size of the discussed effects, we consider a laser beam propagating along the positive $z$ direction, with $\omega_0=1.55\;\text{eV}$, with peak intensity $I_0=10^{22}\;\text{W/cm$^2$}$ ($\xi_0\approx 48$) \cite{Yanovsky_2008}, and with full-width-half maximum duration of $20\;\text{fs}$ in the intensity. Moreover, we consider an electron initially counterpropagating with respect to the plane wave with energy $\varepsilon_i=20\;\text{MeV}$ in such a way that quantum effects can be neglected ($\eta_0\approx 2\times 10^{-3}$ and $\chi_0\approx 10^{-3}$). In this case, $R_C\Phi\approx 0.08$ and, if we choose $\varphi_0$, $\varphi_{\zeta}$ and $\zeta$  such that $g^2=1$, we obtain that $d\mathcal{E}_{G_1}/d\omega|_{\omega\to 0}\approx 3.6\times 10^{-5}$ and $d\mathcal{E}_{G_2}/d\omega|_{\omega\to 0}\approx 2.7\times 10^{-3}$, which also shows the advantage of employing a two-color Gaussian beam. Since in the infrared limit the function $d\mathcal{E}/d\omega$ tends to a constant, the amount $\Delta\mathcal{E}(\omega_M)$ of energy emitted up to a sufficiently small given angular frequency $\omega_M$ is approximately given by $\Delta\mathcal{E}(\omega_M)\approx \omega_Md\mathcal{E}/d\omega|_{\omega\to 0}$. It is interesting to compare the quantity $\Delta\mathcal{E}(\omega_M)$ with the total energy emitted, which can also evaluated exactly by employing the relativistic Larmor formula in Eq. (\ref{Larmor}) and the exact solution of the LL equation in the Appendix A. The exact final results can be written in the compact and manifestly covariant way
\begin{equation}
\mathcal{E}\approx-\frac{2}{3}e^2\int_{-\infty}^{\infty} dt\,\frac{eF^{\mu\nu}u_{\nu}}{m}\frac{eF_{\mu\lambda}u^{\lambda}}{m}=\frac{2}{3}e^2\eta_0\int_{-\infty}^{\infty}d\varphi\,\tilde{\varepsilon}(\varphi)\frac{\xi_1^2\psi^{\prime\,2}_1(\varphi)+\xi_2^2\psi^{\prime\,2}_2(\varphi)}{h(\varphi)},
\end{equation}
where the electron energy $\tilde{\varepsilon}(\varphi)$ is obtained from the analytical solution of the LL equation by neglecting there the terms due to the derivative term in the LL equation (see the Appendix A). A numerical evaluation of this integral for the one- and two-color Gaussian beams considered above gives $\mathcal{E}_{G_1}\approx\mathcal{E}_{G_2}\approx 1\;\text{MeV}$. These values are in excellent agreement with the differences $\varepsilon_i-\varepsilon_{G_1,f}\approx\varepsilon_i-\varepsilon_{G_2,f}\approx 1\;\text{MeV}$ as it results from the conservation of energy. In fact, one can ascertain analytically (see the Appendix A) and numerically,  that the remaining terms in the overall energy conservation equation are negligibly small in both cases. Concerning the quantities $\Delta\mathcal{E}_{G_1}(\omega_M)$ and $\Delta\mathcal{E}_{G_2}(\omega_M)$, we can choose $\omega_M=0.2\,\eta_0\varepsilon_i/\xi_0^2\Delta\varphi\approx 0.02\;\text{eV}$ in our units (see the Appendix B), such that $\Delta\mathcal{E}_{G_1}(\omega_M)\approx 7\times 10^{-8}\;\text{eV}$ and $\Delta\mathcal{E}_{G_2}(\omega_M)\approx 9\times 10^{-6}\;\text{eV}$. As expected, most of the energy is emitted at frequencies much higher than $\omega_M$ but the important result here is that the emission spectrum does not vanish in the limit $\omega\to 0$ as if one neglects radiation reaction.

We conclude by mentioning the consequences of the above results in relation to the underlying and more fundamental theory of QED and, for the sake of convenience, we reintroduce the constants $\hbar$ and $c$. We recall that the study of the infrared behavior of QED and, in particular, of the so-called ``infrared divergence'' goes back to the well-known Bloch-Nordsieck result \cite{Bloch_1937}: the logarithmic infrared divergence in the total number of emitted photons in an arbitrary QED process cancels out once one takes into account self-consistently quantum radiative corrections and the finite resolution of photon detectors (see, e.g., \cite{Jauch_b_1976}). The analysis of the origin of infrared divergences in QED is still active \cite{Lavelle_2006,Kitamoto_2012} especially when processes occur in the presence of a background electromagnetic field \cite{Akhmedov_2009}, in particular of a plane wave \cite{Dinu_2012,Ilderton_2013b}. Now, by formally dividing Eq. (\ref{dE_do}) by $\hbar\omega$ (before taking the limit $\omega\to 0$) and introducing a small fictitious photon mass $\mu$\footnote{Alternatively dimensional regularization can also be employed.}, one obtains that the number of photons $\mathcal{N}(\omega_M;\mu)=\int_{\mu c^2/\hbar}^{\omega_M}d\omega\,(\hbar\omega)^{-1}d\mathcal{E}/d\omega$ emitted with energy $\mu c^2\le\hbar\omega\le\hbar\omega_M$ logarithmically diverges in the limit $\mu\to 0$ \cite{Peskin_b_1995}. This is not contradictory within the classical theory where the number of emitted photons has no physical meaning. Now, on the one hand, the spectrum in Eq. (\ref{dE_dodO}) (once the constant $c$ is appropriately reintroduced) divided by $\hbar\omega$ coincides with the average number of emitted photons calculated quantum mechanically in the classical limit when the recoils of all emitted photons are negligible \cite{Glauber_1951}. On the other hand, the LL equation is classically equivalent to the LAD equation, which has been derived from strong-field QED in \cite{Moniz_1977,Johnson_2002} (see also \cite{Johnson_2000}). The wording ``classically equivalent'' has to be intended as the predictions of the two equations differ by effects scaling with the (classical) parameters $\alpha\eta_0$ and $\alpha\chi_0$\footnote{Recall that $\chi_0=\eta_0\xi_0$ and that in cgs units $\alpha=e^2/\hbar c$ and $\eta_0=\hbar(k_0p)/m^2c^2$, whereas $\xi_0$ is a classical parameter.}, with $\alpha=e^2/\hbar c\approx 1/137$ being the fine-structure constant, which are much smaller than the quantum parameters $\eta_0$ and $\chi_0$ (see Appendix C for additional details). Thus, one can conclude that in the infrared limit of smaller and smaller values of $\omega_M$ the quantity 
\begin{equation}
\mathcal{N}(\omega_M;\mu)=\frac{2}{\pi}\alpha\log\left(\frac{\hbar\omega_M}{\mu c^2}\right)\left[\frac{\rho}{\sqrt{\rho^2-1}}\log\big(\rho+\sqrt{\rho^2-1}\big)-1\right], 
\end{equation}
with $\rho$ given by Eq. (\ref{rho}) provides the classical limit of the corresponding number of emitted photons calculated within strong-field QED in a plane wave, in the sense specified below. In fact, as it is clear from the derivation of the LL equation from the LAD equation \cite{Landau_b_2_1975} (see also Appendix C), the quantity $\mathcal{N}(\omega_M;\mu)$ is exact in the classical (and potentially large) parameter $R_C$ but undergoes additional classical corrections, which scale as $\alpha\eta_0$ and $\alpha\chi_0$ and are, therefore, sub-leading with respect to quantum corrections \footnote{This analysis ignores the unphysical solutions of the LAD equation, which feature a non-perturbative dependence $\sim \exp{(s/\tau)}$ on the parameter $\tau=(2/3)e^2/mc^3$ \cite{Bhabha_1946}.}. These additional corrections are expected to scale with the two classical parameters $\alpha\eta_0$ and $\alpha\chi_0$ and are neglected in the whole above analysis. Indeed, these are about two orders of magnitude smaller than already ignored quantum corrections scaling with the parameters $\eta_0$ and $\chi_0$. In principle, a full quantum calculation should reproduce not only the leading-order classical corrections scaling as $R_C$ and evaluated above but also the additional classical corrections  scaling as $\alpha\eta_0$ and $\alpha\chi_0$, which are sub-leading with respect to the quantum ones. However, in order to obtain solely the classical limit found above an appropriate sub-set among all possible quantum processes may turn out to be sufficient.

\section{Conclusions}

In conclusion, we have derived analytically the infrared limit of nonlinear Thomson scattering, which is valid to all orders in the classical parameter $R_C=\alpha\chi_0\xi_0$, to leading order in the classical parameters $\alpha\eta_0$ and $\alpha\chi_0$, and for an arbitrary plane wave. The result shows that classical radiative corrections qualitatively and quantitatively alter the infrared behavior with respect to the leading-order Lorentz dynamics. On the one hand, these results can be employed in principle as signatures of classical radiation reaction. On the other hand, they represent the classical limit of the corresponding infrared behavior of strong-field QED in a plane wave, exact in $R_C$ and to leading order in $\alpha\eta_0$ and $\alpha\chi_0$. As a byproduct, we have also provided the analytical expression of the total energy emitted by an electron in an arbitrary plane wave by taking into account radiation reaction to all orders in $R_C$.

\appendix

\section*{Appendix A}
\setcounter{equation}{0}
\renewcommand{\theequation}{A.\arabic{equation}}

In the present appendix we report the analytical solution of the Landau-Lifshitz (LL) equation \cite{Landau_b_2_1975} in a plane wave as given in \cite{Di_Piazza_2008_a} and the resulting overall energy-momentum conservation equation for the electron. For the sake of completeness, we also report the LL equation in an external electromagnetic field $F^{\mu\nu}=F^{\mu\nu}(x)$ \cite{Landau_b_2_1975}
\begin{equation}
\label{iLL_eq}
\begin{split}
m\frac{d u^{\mu}}{ds}=&eF^{\mu\nu}u_{\nu}+\frac{2}{3}e^2\left[\frac{e}{m}(\partial_{\alpha}F^{\mu\nu})u^{\alpha}u_{\nu}\right.\\
&\left.+\frac{e^2}{m^2}F^{\mu\nu}F_{\nu\alpha}u^{\alpha}+\frac{e^2}{m^2}(F^{\alpha\nu}u_{\nu})(F_{\alpha\lambda}u^{\lambda})u^{\mu}\right],
\end{split}
\end{equation}
where the notation and the units are the same as in the main text. In the case of the plane wave as defined in the main text, it is convenient to introduce the four-wave-vector $k_0^{\mu}=\omega_0n_0^{\mu}$, where $\omega_0$ is the central angular frequency (or, more in general, an arbitrary frequency scale describing the time dependence of the plane wave), and the laser phase $\varphi=(k_0x)$. Analogously to the case of the Lorentz equation, it is natural to use $\varphi$ as the independent variable to solve the LL equation. By choosing the initial condition of the electron four-momentum as $\lim_{\varphi\to-\infty}p^{\mu}(\varphi)=p_i^{\mu}$, the four-momentum $p^{\mu}(\varphi)$ reads \cite{Di_Piazza_2008_a}
\begin{equation}
\label{iSol_A}
\begin{split}
p^{\mu}(\varphi)=&\frac{1}{h(\varphi)}\left\{p_i^{\mu}+\frac{1}{2\eta_0}[h^2(\varphi)-1]k_0^{\mu}+\frac{1}{\eta_0}\left[\mathcal{I}_1(\varphi)\frac{eF_1^{\mu\nu}}{m^2}+\mathcal{I}_2(\varphi)\frac{eF_2^{\mu\nu}}{m^2}\right]p_{i,\nu}\right.\\
&\left.+\frac{1}{2\eta_0}[\xi_1^2\mathcal{I}^2_1(\varphi)+\xi_2^2\mathcal{I}^2_2(\varphi)]k_0^{\mu}\right\},
\end{split}
\end{equation}
where we have introduced the constant quantities $\eta_0=(k_0p_i)/m^2$, $F_j^{\mu\nu}=k_0^{\mu}a_j^{\nu}-k_0^{\nu}a_j^{\mu}$, and $\xi_j=|e\bm{a}_j|/m$, with $j=1,2$, and the functions
\begin{equation}
h(\varphi)= 1+\frac{2}{3}e^2\eta_0\int_{-\infty}^{\varphi}d\tilde{\varphi}[\xi_1^2\psi^{\prime\,2}_1(\tilde{\varphi})+\xi_2^2\psi^{\prime\,2}_2(\tilde{\varphi})]
\end{equation}
and
\begin{equation}
\mathcal{I}_j(\varphi)=\psi_j(\varphi)h(\varphi)+\frac{2}{3}e^2\eta_0\psi'_j(\varphi)-\frac{2}{3}e^2\eta_0\int_{-\infty}^{\varphi}d\tilde{\varphi}\,\psi_j(\tilde{\varphi})[\xi_1^2\psi^{\prime\,2}_1(\tilde{\varphi})+\xi_2^2\psi^{\prime\,2}_2(\tilde{\varphi})].
\end{equation}
Note that the solution of the Lorentz equation is formally obtained by setting $e^2=0$ in the functions $h(\varphi)$ and $\mathcal{I}_j(\varphi)$ (see, e.g., \cite{Landau_b_2_1975}). Also, if one assumes that $|\psi'_j(\varphi)|\sim |\psi_j(\varphi)|$, the term proportional to $\psi'_j(\varphi)$ in $\mathcal{I}_j(\varphi)$ can be neglected according to Landau and Lifshitz reduction of order \cite{Landau_b_2_1975}. This can also be seen directly from the LL equation because $eF^{\mu\nu}u_{\nu}+(2/3)e^2(e/m)(\partial_{\alpha}F^{\mu\nu})u^{\alpha}u_{\nu}=e[F^{\mu\nu}+(2/3)e^2\eta_0 F^{\prime\,\mu\nu}/h(\phi)]u_{\nu}$, which shows that the second term is about $\alpha\eta_0$ times smaller than the first one (see \cite{Landau_b_2_1975}). Having neglected this term, the overall energy-momentum conservation law is obtained by integrating the remaining terms of the LL equation with respect to the proper time $s$, by performing the change of variable from $s$ to $\varphi$ [$ds=h(\varphi)d\varphi/m\eta_0$], and by replacing everywhere the exact solution. The result can be written in the form $p_f^{\mu}-p_i^{\mu}=W^{\mu}+P^{\mu}+R^{\mu}$, where
\begin{equation}
W^{\mu}=\frac{1}{3}e^2k_0^{\mu}\int_{-\infty}^{\infty}d\varphi\,[\xi_1^2\psi^2_1(\varphi)+\xi_2^2\psi^2_2(\varphi)][\xi_1^2\psi^{\prime\,2}_1(\varphi)+\xi_2^2\psi^{\prime\,2}_2(\varphi)]
\end{equation}
arises from the Lorentz-four-force in the LL equation and is the energy-momentum transferred by the plane wave to the electron as it moves inside the plane wave (this quantity also vanishes if radiation-reaction effects are ignored), where
\begin{equation}
P^{\mu}=\frac{2}{3}e^2k_0^{\mu}\int_{-\infty}^{\infty}d\varphi\,[\xi_1^2\psi^{\prime\,2}_1(\varphi)+\xi_2^2\psi^{\prime\,2}_2(\varphi)]
\end{equation}
arises from the first term in the second line of the LL equation [see Eq. (\ref{iLL_eq})] and is the energy-momentum transferred from the plane wave to the electron concurrently with the emission of radiation, and where
\begin{equation}
R^{\mu}=-\frac{2}{3}e^2\eta_0\int_{-\infty}^{\infty}d\varphi\,\tilde{p}^{\mu}(\varphi)\frac{\xi_1^2\psi^{\prime\,2}_1(\varphi)+\xi_2^2\psi^{\prime\,2}_2(\varphi)}{h(\varphi)}
\end{equation}
arises from the second term in the second line of the LL equation [the Larmor term, see Eq. (\ref{iLL_eq})] and is the energy-momentum radiated by the electron. Here, $\tilde{p}^{\mu}(\varphi)$ is obtained from Eq. (\ref{iSol_A}) by neglecting the terms in $\mathcal{I}_j(\varphi)$ due to the derivative term in the LL equation.

\section*{Appendix B}
\setcounter{equation}{0}
\renewcommand{\theequation}{B.\arabic{equation}}

In this appendix we provide a suitable expression for the upper limit $\omega_M$ of the emitted angular frequencies $\omega$ such that the phase $\Psi=\omega_M(t-\bm{n}\cdot\bm{r})$ can be neglected in Eq. (\ref{dE_dodO}). Since we need only an order-of-magnitude estimate of $\omega_M$, for the sake of simplicity: 1) we neglect radiation-reaction effects; 2) we assume, as it is also usually the case in experiments \cite{Cole_2018,Poder_2017}, that the electron is initially counterpropagating with respect to the laser field with energy $\varepsilon_i\gg m\xi_0\gg m$; 3) we look at emission directions $\bm{n}$ where most of the radiation is emitted such that if $\bm{n}\cdot\bm{\beta}=\beta\cos\theta$, then $\theta\lesssim\theta^*=\xi_0/\gamma_i\ll 1$ (see, e.g., \cite{Baier_b_1998}). Without loss of generality for the estimation of $\omega_M$, we can also assume that $\bm{r}(0)=\bm{0}$, such that
\begin{equation}
\label{Psi}
\Psi\lesssim\frac{\omega_M}{m\eta_0}\int_0^{\varphi}d\tilde{\varphi}\,\gamma_0(\tilde{\varphi})\left[1-\beta_0(\tilde{\varphi})+\frac{\theta^{*\,2}}{2}\right],
\end{equation}
where $\beta_0(\varphi)$ and $\gamma_0(\varphi)$ are the phase dependent modulus of the electron velocity and its Lorentz factor according to the Lorentz equation (see the Appendix A). Now, the pre-exponential function in Eq. (\ref{dE_dodO}) is different from zero only while the electron is accelerated. Thus, if $\Delta\varphi$ denotes a measure of the plane wave pulse duration, we can say that $\varphi\lesssim\Delta\varphi$ in Eq. (\ref{Psi}) such that
\begin{equation}
\Psi\lesssim\frac{\omega_M}{2\varepsilon_i\eta_0}(1+\theta^{*\,2}\gamma_i^2)\Delta\varphi.
\end{equation}
Recall, in fact, that under the above assumptions it is $\gamma_0(\varphi)\approx \gamma_i\gg 1$ (see the Appendix A). In conclusion, for the sake of estimate we can set $\omega_M=0.2\,\eta_0\varepsilon_i/\xi_0^2\Delta\varphi$.

\section*{Appendix C}
\setcounter{equation}{0}
\renewcommand{\theequation}{C.\arabic{equation}}

In the present appendix, we would like to show explicitly that the LL equation (\ref{iLL_eq}) differs from the LAD equation \cite{Abraham_b_1905,Lorentz_b_1909,Dirac_1938}
\begin{equation}
\label{LAD}
m\frac{du^{\mu}}{ds}=eF^{\mu\nu}u_{\nu}+\frac{2}{3}e^2\left(\frac{d^2u^{\mu}}{ds^2}+\frac{du^{\nu}}{ds}\frac{du_{\nu}}{ds}u^{\mu}\right)
\end{equation}
only by terms being in order of magnitude about $1/\alpha\approx 137$ times smaller than quantum effects. We recall that we employ units with $\hbar=c=4\pi\epsilon_0=1$, such that $\alpha=e^2$. It is clear that, depending on the specific space-time configuration of the external electromagnetic field, a more detailed analysis may be required to draw quantitative conclusions and here we limit ourselves to study the general scaling of the various terms in the LL and the LAD equations. Now, it is first convenient to introduce the dimensionless quantities
\begin{align}
\kappa=\frac{e}{|e|}, && \zeta^{\mu}=\frac{3}{2}\frac{x^{\mu}}{r_0}, && \sigma=\frac{3}{2}\frac{s}{r_0}, && \Phi^{\mu\nu}(x)=\frac{2}{3}\frac{F^{\mu\nu}(x)}{F_0}.
\end{align}
Here, we have introduced the quantities $r_0=e^2/m=\alpha\lambda_C$ (classical electron radius) and $F_0=m^2/|e|^3=F_{cr}/\alpha$ (critical field of classical electrodynamics), where $\lambda_C=1/m$ (Compton wavelength) and $F_{cr}=m^2/|e|$ (critical field of QED) are the typical length scale and field scale of QED \cite{Landau_b_4_1982}. 

By multiply the LAD equation by $(2/3)e^2/m^2$, it can be written in the convenient form 
\begin{equation}
\label{LAD_0}
\frac{du^{\mu}}{d\sigma}=\kappa\Phi^{\mu\nu}u_{\nu}+\frac{d^2u^{\mu}}{d\sigma^2}+\frac{du^{\nu}}{d\sigma}\frac{du_{\nu}}{d\sigma}u^{\mu}.
\end{equation}
Now, the original LL equation is obtained by replacing the four-acceleration in the radiation-reaction force with the Lorentz force divided by $m$ up to terms of the order of $e^4$. In order to obtain the LL equation up to the order $e^6$, we need to expand Eq. (\ref{LAD_0}) up to the order $e^8$ having in mind that $u^{\mu}=O(e^0)$, $d/d\sigma=O(e^2)$, and $\Phi^{\mu\nu}(x)=O(e^3)$. It is laborious but straightforward to show that the LL equation up to the order $e^6$ can be written in the convenient form
\begin{equation}
\label{LL_6}
\begin{split}
\frac{du^{\mu}}{d\sigma}=&\kappa\Phi^{\mu\nu}u_{\nu}+\kappa[\nabla_{\lambda}(\Phi^{\mu\nu})u_{\nu}+\nabla_{\lambda}\nabla_{\rho}(\Phi^{\mu\nu})u^{\rho}u_{\nu}]u^{\lambda}\\
&+[\Phi^{\mu\lambda}\Phi_{\lambda\rho}+2\nabla_{\lambda}(\Phi^{\mu\nu}\Phi_{\nu\rho})+\nabla^{\lambda}(\Phi^{\mu\nu})\Phi_{\lambda\rho}u_{\nu}]u^{\rho}\\
&+\Phi^{\nu\rho}u_{\rho}[\Phi_{\nu\lambda}+4\nabla_{\beta}(\Phi_{\nu\lambda})u^{\beta}]u^{\lambda}u^{\mu},
\end{split}
\end{equation}
where the symbol $\nabla_{\mu}$ indicates the derivative with respect to the dimensionless coordinates $\zeta^{\mu}$, such that $\nabla_{\mu}=O(e^2)$. Now, the first term in each square bracket corresponds to the LL equation and all others are the higher-order corrections, of the order of either $e^5$ or $e^6$. By indicating as $\chi$ and $\eta$ the parameters reducing to $\chi_0$ and to $\eta_0$ in the case of a plane wave, it is clear that $\sqrt{-\Phi^{\mu\lambda}(x)u_{\lambda}\Phi_{\mu\nu}(x)u^{\nu}}\sim\alpha\chi$ and $u^{\mu}\nabla_{\mu}\sim \alpha\eta$, such that the higher-order terms are always either $\alpha\chi$ or $\alpha\eta$ times smaller than the leading-order terms included in the original LL equation. We conclude by making two remarks. Firstly, the form (\ref{LAD_0}) of the LAD equation and the scaling of each term in it clearly shows that the above conclusion will be true also at higher orders in $e$. In this respect, it is worth noticing that, in constructing higher-order terms, the four-vector $u^{\mu}$ in the last term of the LAD equation can only be contracted either with $\Phi_{\mu\nu}(x)$ or with $\nabla_{\mu}$, resulting into corrections of the order of either $\alpha\chi$ or $\alpha\eta$, respectively. Secondly, we draw attention to the last term $(\Phi^{\nu\rho}u_{\rho}\Phi_{\nu\lambda}u^{\lambda})u^{\mu}$ in Eq. (\ref{LL_6}) proportional to $u^{\mu}$. Due to the extra Lorentz factor arising because of the un-contracted four-vector $u^{\mu}$, the classical parameter $R_C=\alpha\chi_0\xi_0$ appears in the analytical solution of the LL equation in a plane wave (see the Appendix A), which can be much larger than $\alpha\chi_0$. However, the correction $4[\Phi^{\nu\rho}u_{\rho}\nabla_{\beta}(\Phi_{\nu\lambda})u^{\beta}u^{\lambda}]u^{\mu}$ to this term in Eq. (\ref{LL_6}) is also proportional to $u^{\mu}$, such that the effects arising from this correction will always be $\alpha\eta$ times smaller than the effects scaling with $R_C$ and then negligible with respect to the quantum corrections to the latter effects.

\color{black}

\section*{Acknowledgments}
I am grateful to O. Skoromnik for an insightful discussion on quantum infrared divergences.




\end{document}